# Steady state effects introduced by local relaxation modes on J-driven DNP-enhanced NMR

Maria Grazia Concilio* and Lucio Frydman

*Department of Chemical and Biological Physics, Weizmann Institute of Science, Rehovot, Israel*

**Abstract**

One of solution-state Nuclear Magnetic Resonance (NMR)'s main weaknesses, is its relative insensitivity. J-driven Dynamic Nuclear Polarization (JDNP) was recently proposed for enhancing solution-state NMR's sensitivity, by bypassing the limitations faced by conventional Overhauser DNP (ODNP), at the high magnetic fields where most analytical research is performed. By relying on biradicals with inter-electron exchange couplings $J_{ex}$ on the order of the electron Larmor frequency $\omega_E$, JDNP was predicted to introduce a transient enhancement in NMR's nuclear polarization at high magnetic fields, and for a wide range of rotational correlation times of medium-sized molecules in conventional solvents. This communication revisits the JDNP proposal, including additional effects and conditions that were not considered in the original treatment. These include relaxation mechanisms arising from local vibrational modes that often dominate electron relaxation in organic radicals, as well as the possibility of using biradicals with $J_{ex}$ of the order of the nuclear Larmor frequency $\omega_N$ as potential polarizing agents. The presence of these new relaxation effects lead to variations in the JDNP polarization mechanism originally proposed, and indicate that triplet-to-singlet cross-relaxation processes may lead to a nuclear polarization enhancement that persists even at steady states. The physics and potential limitations of the ensuing theoretical derivations, are briefly discussed.

* maria-grazia.concilio@weizmann.ac.il

## 1. Introduction

Nuclear Magnetic Resonance (NMR) is one of the most versatile forms of spectroscopy, conveying structural and dynamical information with minimal invasiveness. Even further applications could emerge if it was not for the limited sensitivity of NMR – particularly when executed on room temperature solutions and at high magnetic fields (7 T - 23 T), where most analytical and biophysical studies performed. The sensitivity of solution state NMR can be enhanced by Overhauser DNP (ODNP) [1-3]; however, when driven by dipolar relaxation mechanisms, ODNP only works efficiently at relatively low magnetic fields [4-7]. High-field ODNP experiments by contrast are usually limited, as they need to be aided by Fermi contact couplings of the kind which arises for certain nuclei such as $^{31}$P [8, 9], $^{19}$F [10, 11] and $^{13}$C [6, 12-14]. Significant scalar-driven DNP enhancements can then arise, but this requires an electron delocalization that only arises in specific radical/solvent combinations. [6, 11, 15-18]. However, for the more general and highly relevant case, involving $^{1}$H nuclei interacting with electron radicals solely through inter-molecular dipolar hyperfine couplings, ODNP efficiency decays rapidly with the magnetic field, $B_0$. [3, 19]

*J*-driven Dynamic Nuclear Polarization (JDNP) [20] is a recently described theoretical proposal, aiming to enhance NMR's sensitivity in solution state at any magnetic field, solely relying on inter-molecular dipolar hyperfine couplings. As ODNP, JDNP can be conceivably performed either with a microwave irradiation at the electron Larmor frequency or by shuttling the sample between higher and lower magnetic fields, with one of these fields serving to achieve hyperpolarization and the other NMR observation. [21, 22] Unlike ODNP, which relies on stable mono-radicals, JDNP relies on stable biradicals with the inter-electron exchange coupling $J_{ex}$ in the range of the electron Larmor frequency $\omega_E$. In such case, Redfield's relaxation theory [23] predicts that at the $J_{ex} \approx \omega_E$ JDNP condition, a difference between the self-relaxation rates of the two-electron singlet and triplet states which are dipolar hyperfine coupled to the α or β nuclear states will arise. This leads to a transient imbalance between the nuclear spin populations, and consequently a transient nuclear polarization build-up.

The current work describes the effects of non-Redfield, field-independent local vibrational modes arising from the mixing between spin and orbital angular momentum, on the JDNP enhancement. Relaxation arising from local vibrational modes driven by spin-orbit coupling, [24-27] often dominate the electron relaxation rates of organic radicals [24, 25, 27, 28]. These effects are shown to lead to steady-state NMR signal enhancements if moderate, and to the suppression of the JDNP phenomena if overtly dominant. For completion this study also considers the possibility of using biradicals with $J_{ex}$ of the order of the nuclear Larmor frequency $\omega_N$, for enhancing the sensitivity of solution state NMR at high magnetic fields.

## 2. Theoretical methodologies

This study's steady state numerical simulations were performed using the Spinach software package [29] based on a laboratory frame Hamiltonian, using the Fokker-Planck formalism described in [30]. Numerical simulations took account of all self- and cross-relaxation term within the Bloch-Redfield-Wangsness relaxation theory, [23, 31] as well as scalar relaxation of the first kind [32] and relaxation from a local vibrational mode. Scalar relaxation of the first kind arises from the conformational mobility that modulates the exchange coupling. A "pessimistic" modulation depth of 3 GHz was assumed for this, with a conformational mobility correlation time of 1 picosecond. Still, it was found that these parameters have negligible effects on the JDNP enhancement, since in the $\omega_{E1} - \omega_{E2} \to 0$ scenario here considered – where $\omega_{E1}$ and $\omega_{E2}$ correspond to the Larmor frequencies of the two

electrons – the exchange coupling commutes with the Zeeman interaction leading to no additional relaxation effects.

Relaxation from a local vibrational mode was included as an additional diagonal term in the Redfield relaxation superoperator, applied to the electron longitudinal and transverse states. In the case of trityls, local mode relaxation presumably arises due the stretching of the C–S bonds in radical structures and its rate constant is about $6 \times 10^4$ Hz; [24-27] this was the value assumed throughout most of our simulations.

The spin system is composed by two electrons and one proton that interact solely through dipolar anisotropic hyperfine couplings. Distances between the two unpaired spin-1/2 electrons (belonging to the radical) and the spin-1/2 proton (assumed to belong to a stationary solvent molecule) were set to 5.5 Å and 13.2 Å respectively (in the Table 1 the actual proton and electron Cartesian coordinates are reported). The electron $g$-tensors were set axially symmetric, with eigenvalues taken from those found in trityls. [33, 34] The $g$-tensors were set non-colinear. As discussed previously, [22] in the case of axially symmetric $g$-tensors, rotations perpendicular to the axis along the linker connecting the two monomeric units in a symmetric biradical, could lead to variations that would subtract efficiency from JDNP. These bending angles would be small for short linkers, but could range between ≈0° and 40° for long linkers containing multiple para-phenylene or acetylene units. [35] Short and long linkers could lead in turn to $J_{ex}$ values ranging from a few 100s MHz, to up to 300 GHz [36, 37]. For the sake of simplicity, a $\beta$ = 20° bending angle was adopted throughout out calculations; a more comprehensive analysis on the effect of these distortions is presented in Supplementary Information A. All the simulation parameters are summarized in Table 1.

**Table 1:** Biradical / proton spin system parameters used in this paper's simulations.

| Parameter | Spin system |
|---|---|
| $^1$H chemical shift tensor, ppm | [5 5 5] |
| Electron 1 $g$-tensor[1] eigenvalues, [xx yy zz] / Bohr magneton | [2.0032 2.0032 2.0026] |
| Electron 1 $g$-tensor, ZYZ active Euler angles / rad | [0.0 0.0 0.0] |
| Electron 2 $g$-tensor[1] eigenvalues, [xx yy zz] / Bohr magneton | [2.0032 2.0032 2.0026] |
| Electron 2 $g$-tensor, ZYZ active Euler angles / rad | [0.0 π/8 0.0] |
| $^1$H coordinates [x y z] / Å | [0.0 5.0 5.0] |
| Electron 1 and electron 2 coordinates, [x y z] / Å | [0 0 -7.20] and [0 0 7.20] |
| Rotational correlation time[2] $\tau_C$ / ps | 800 |
| Scalar relaxation modulation depth / GHz | 3 |
| Scalar relaxation modulation time / ps | 1 |
| Local mode relaxation / Hz | $6 \times 10^4$ |
| Temperature / K | 298 |

[1] Simulations used Zeeman interaction tensors that were made axially symmetric along the main molecular axis (corresponding for instance, to a linker connecting two trityl units).
[2] Rotational correlation time of the biradical/proton dipolar hyperfine coupled triad.

### 3. Results

#### 3.1 Features of the steady state JDNP

The transient nuclear polarization build-up obtained in the previously mentioned JDNP, based only on the Redfield relaxation superoperator model, [20] becomes stable at a non-zero steady state when relaxation from local vibrational modes are included in the relaxation superoperator. [29, 31] Figure 1 shows this effect, with an increase of the DNP enhancement with an increase of the microwave nutation power. As can be seen, when the microwave nutation rates become comparable or higher than the magnitude of the relaxation rate arising from the local vibrational mode (which in the present model dominates the electron relaxation rate), a significant steady state enhancement of the nuclear polarization can be achieved. At 9.4 T, where microwave nutation powers of up to about 4 MHz have been reported, [38] high enhancements will arise even if local mode relaxation rates are allowed to increase to 1 MHz –a typical value observed for organic monoradicals in fluid solutions. [26, 27, 39]

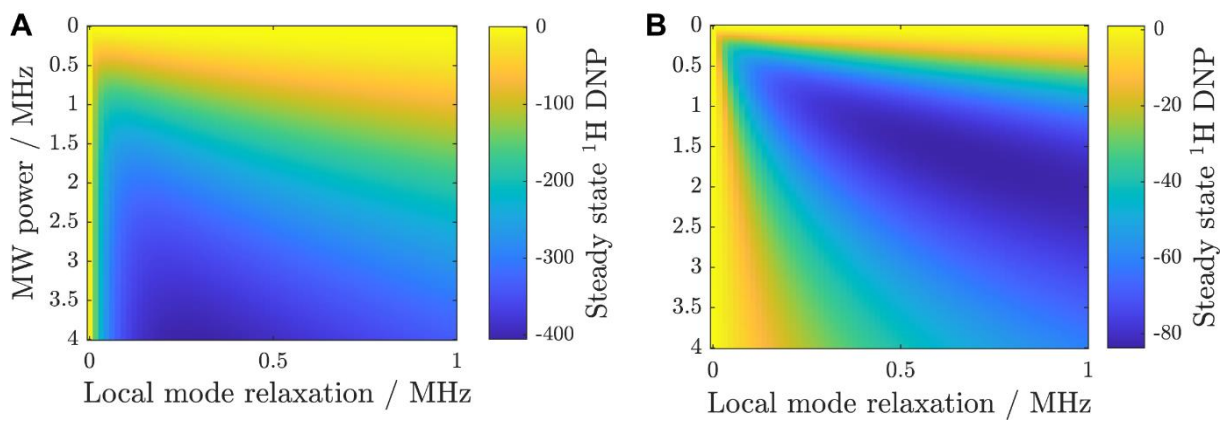

**Figure 1:** Spinach [29] simulated DNP enhancements obtained at 9.4 T upon continuous microwave irradiation as a function of the microwave power ($\gamma B_1$ nutation frequencies) and of the magnitude local mode relaxation rate, applied to the diagonal terms in the Redfield relaxation superoperator. Conditions explored $J_{ex}=\omega_E+\omega_N$ in (A) and $J_{ex}=\omega_N$ in (B). In this and other figures shown below, the DNP enhancements denote the achieved nuclear polarization, normalized by its Boltzmann counterpart at the same temperature and field. The simulation parameters are given in Table 1.

Figure 2 shows how the nuclear magnetization enhancement depends on the isotropic exchange coupling between the electrons, as a function of magnetic field $B_0$. A continuous on-resonance microwave irradiation at the electron Larmor frequency of the biradicals was assumed, and steady state nuclear polarizations were calculated. A significant steady state enhancement is predicted at the $J_{ex} = \omega_E+\omega_N$ condition; a more modest but still sizable enhancement is also expected at the condition $J_{ex} = \omega_N$. Notice that the JDNP enhancement's "width" is ($\omega_E+\omega_N$) ± 2 GHz at any field for the $J_{ex} = \omega_E+\omega_N$ condition, while it depends on the magnetic field for the $J_{ex} = \omega_N$ case. For example at 9.4 T, JDNP enhancement is achieved within ca. $\omega_N$ ± 200 MHz.

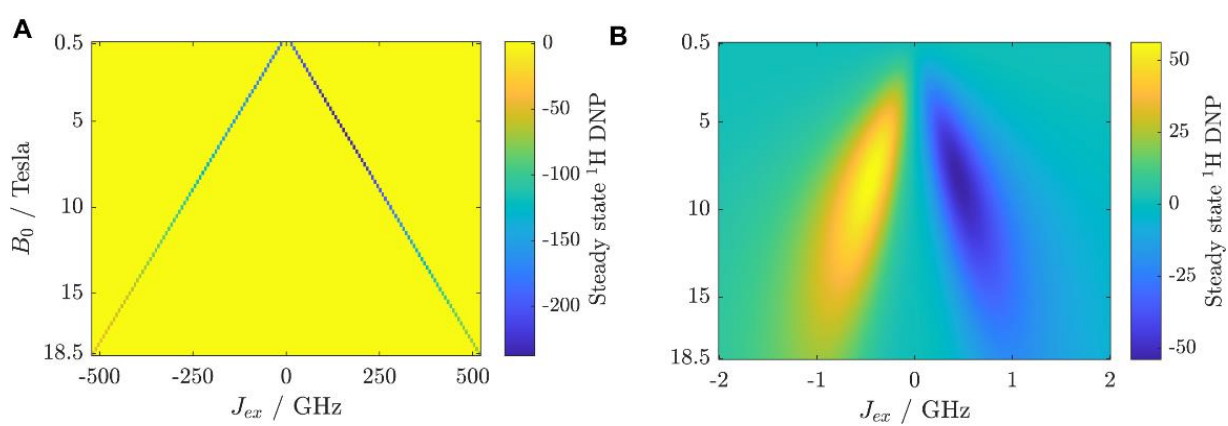

**Figure 2:** Simulated DNP enhancements arising at the steady state, assuming a continuous microwave irradiation of a biradical with a nutation frequency of 500 kHz, as a function of $J_{ex}$ and of $B_0$. The $J_{ex}$ was changed in between –500 and +500 GHz range in (A) and between -2000 and +2000 MHz range in (B). Additional simulation parameters are given in Table 1.

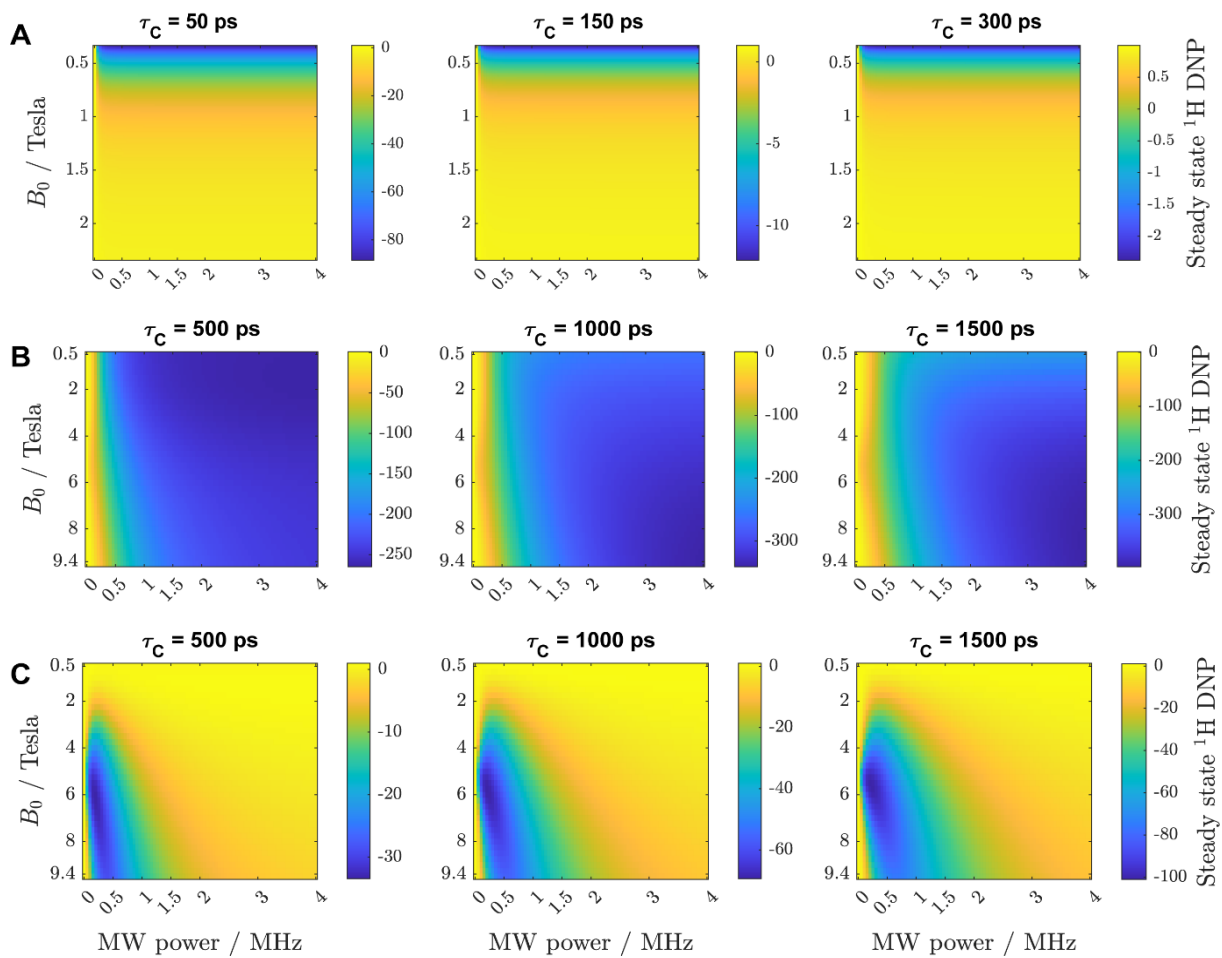

**Figure 3:** Simulated DNP enhancements expected upon continuous on-resonance microwave irradiation at the electron Larmor frequency of the biradical, as a function of the magnetic field, of the microwave nutation power, and of the correlation time $\tau_C$ in (A) ODNP, (B) JDNP with $J_{ex} = \omega_E + \omega_N$ and (C) JDNP with $J_{ex}=\omega_N$. Notice that each plot has its own colorbar scale. The simulation parameters are given in the Table 1.

Figure 3 shows the performance predicted for the JDNP effects as compared with the conventional Overhauser DNP at different magnetic fields, microwave nutation powers and rotational correlation times, $\tau_C$. As expected, the ODNP enhancement, based solely on dipolar hyperfine couplings, is strong when the magnetic field is about 0.5 T, but it decays to negligible values at higher magnetic fields ≥ 1 T, when the rotational correlation time is ≥ 150 ps. The JDNP remains strong at any magnetic field and increases with the $\tau_C$, at both the $J_{ex}=\omega_E+\omega_N$ and $J_{ex}=\omega_N$ conditions. In the case of $J_{ex}=\omega_E+\omega_N$, the DNP enhancement increases with the microwave nutation power, while it decreases with the power in the case of $J_{ex}=\omega_N$, after having reached a maximum at about 200 kHz at 5 T, for this set of simulation parameters.

### 3.2. The physics of the steady state JDNP

The electron relaxation and spin dynamics were examined in this study using the singlet and triplet basis sets. This is justified by the $J_{ex} \gg \omega_{E1} - \omega_{E2}$ scenario, where the electron Zeeman eigenstates are no longer eigenfunctions of the spin Hamiltonian; it is therefore convenient to treat the Hamiltonian in the singlet/triplet electron basis set. Further, as the two electrons are considered localized in different radical monomers, the zero-field splitting is expected to be negligible with respect to the Zeeman interaction, and it was not here included in the Hamiltonians. [40] By contrast, inter – electron dipolar interactions were included in both the spin Hamiltonian and relaxation superoperator in all the simulations; notice that the modulation of these inter-electron dipolar interactions does not contribute to the JDNP mechanism but only to the electron T$_1$ and T$_2$. [20]

The physics of the JDNP can then be described using three-spin population operators, corresponding to the α and β nuclear components of the two-electron singlet and triplet states ($\hat{T}_+$, $\hat{T}_0$, $\hat{T}_-$ and $\hat{S}_0$). [41, 42] An alternative description of JDNP's physics can be done using Cartesian operators, by examining the cross-relaxation rate $\hat{E}_Z \to \hat{N}_Z$; this is analogous to what is usually done in ODNP, and is discussed in the Supporting Material. The steady state effect introduced in Figures 1-3 can be explained considering the Liouvillian:

$$\hat{\hat{L}} = \hat{\hat{H}} + i\hat{\hat{R}} \tag{1}$$

where $\hat{\hat{R}}$ is the Redfield relaxation superoperator and $\hat{\hat{H}}$ is the spin Hamiltonian for a three-spin system composed by two electrons and one proton:

$$\hat{\hat{H}} = \left(\omega_E + \omega_{off}\right)\left(\hat{E}_{1Z} + \hat{E}_{2Z}\right) - \omega_N \hat{N}_Z + \\ + J_{ex}\left(\hat{E}_{1X}\hat{E}_{2X} + \hat{E}_{1Y}\hat{E}_{2Y} + \hat{E}_{1Z}\hat{E}_{2Z}\right) + \omega_{MW}\left(\hat{E}_{1X} + \hat{E}_{2X}\right) \tag{2}$$

where $\omega_{off} = \omega_E - \omega_e$ is the offset between the radical Larmor frequency and the free electron frequency; $\omega_{MW}$ is the micro-wave irradiation power. Terms describing the secular and pseudo-secular dipolar hyperfine interactions (in the order of MHz) are here neglected when compared to terms describing Zeeman and inter-electron exchange interactions (in the order of GHz) –even if they were accounted for and take part in the JDNP mechanism via the relaxation superoperator (as explained below). Although for simplicity these terms are omitted from the spin Hamiltonian in Eq. 2, the full spin Hamiltonian and its singlet/triplet representation is shown in the supporting material of [20]. The spin Hamiltonian in Eq. (2) corresponds to a matrix that can be diagonalized, yielding eigenvalues and eigenstates. For the nuclear α states these eigenvalues correspond to:

$$E_1 = -\frac{3J_{ex}}{4} + \frac{\omega_N}{2} \to |\hat{S}_{0,\alpha}\rangle \qquad E_2 = \frac{J_{ex}}{4} + \frac{\omega_N}{2} + \sqrt{\omega_{MW}^2 + \left(\omega_{off} + \omega_E\right)^2} \to |\hat{T}_{+,\alpha}\rangle \\ E_3 = \frac{J_{ex}}{4} + \frac{\omega_N}{2} \to |\hat{T}_{0,\alpha}\rangle \qquad E_4 = \frac{J_{ex}}{4} + \frac{\omega_N}{2} - \sqrt{\omega_{MW}^2 + \left(\omega_{off} + \omega_E\right)^2} \to |\hat{T}_{-,\alpha}\rangle \tag{3}$$

While for the β states they correspond to:

$$E_5 = -\frac{3J_{ex}}{4} - \frac{\omega_N}{2} \to |\hat{S}_{0,\beta}\rangle \qquad E_6 = \frac{J_{ex}}{4} - \frac{\omega_N}{2} + \sqrt{\omega_{MW}^2 + \left(\omega_{off} + \omega_E\right)^2} \to |\hat{T}_{+,\beta}\rangle \\ E_7 = \frac{J_{ex}}{4} - \frac{\omega_N}{2} \to |\hat{T}_{0,\beta}\rangle \qquad E_8 = \frac{J_{ex}}{4} - \frac{\omega_N}{2} - \sqrt{\omega_{MW}^2 + \left(\omega_{off} + \omega_E\right)^2} \to |\hat{T}_{-,\beta}\rangle \tag{4}$$

At $t = 0$, when $\omega_{MW} = 0$, the eigenvalues of the α states become equal to:

$$E_1(0) = -\frac{3J_{ex}}{4} + \frac{\omega_N}{2} \to |\hat{S}_{0,\alpha}\rangle \qquad E_2(0) = \frac{J_{ex}}{4} + \frac{\omega_N}{2} + \omega_{off} + \omega_E \to |\hat{T}_{+,\alpha}\rangle \\ E_3(0) = \frac{J_{ex}}{4} + \frac{\omega_N}{2} \to |\hat{T}_{0,\alpha}\rangle \qquad E_4(0) = \frac{J_{ex}}{4} + \frac{\omega_N}{2} + \omega_{off} - \omega_E \to |\hat{T}_{-,\alpha}\rangle \tag{5}$$

and the eigenvalues of the β states become equal to:

$$E_5(0) = -\frac{3J_{ex}}{4} - \frac{\omega_N}{2} \rightarrow |\hat{S}_{0,\beta}\rangle \quad E_6(0) = \frac{J_{ex}}{4} - \frac{\omega_N}{2} + \omega_{off} + \omega_E \rightarrow |\hat{T}_{+,\beta}\rangle$$
$$E_7(0) = \frac{J_{ex}}{4} - \frac{\omega_N}{2} \rightarrow |\hat{T}_{0,\beta}\rangle \quad E_8(0) = \frac{J_{ex}}{4} - \frac{\omega_N}{2} + \omega_{off} - \omega_E \rightarrow |\hat{T}_{-,\beta}\rangle \quad (6)$$

The energy of the eigenstates in Eqs. (5) and (6) can be arranged from low to high; in the case of negative $J_{ex} \approx \omega_E$, the order is:

$$E_4 < E_8 < E_3 < E_7 < E_1 < E_2 < E_5 < E_6 \quad (7)$$

with $E_1 \sim E_2 \sim E_5 \sim E_6$. In the case of positive $J_{ex} \approx \omega_N$, the order is:

$$E_4 < E_8 < E_1 < E_3 < E_5 < E_7 < E_2 < E_6 \quad (8)$$

with $E_1 \sim E_3 \sim E_5 \sim E_7$. In the absence of microwave irradiation, this leads to the thermal equilibrium populations shown in the central panels of Figures 4A and 4B (MW power = 0).

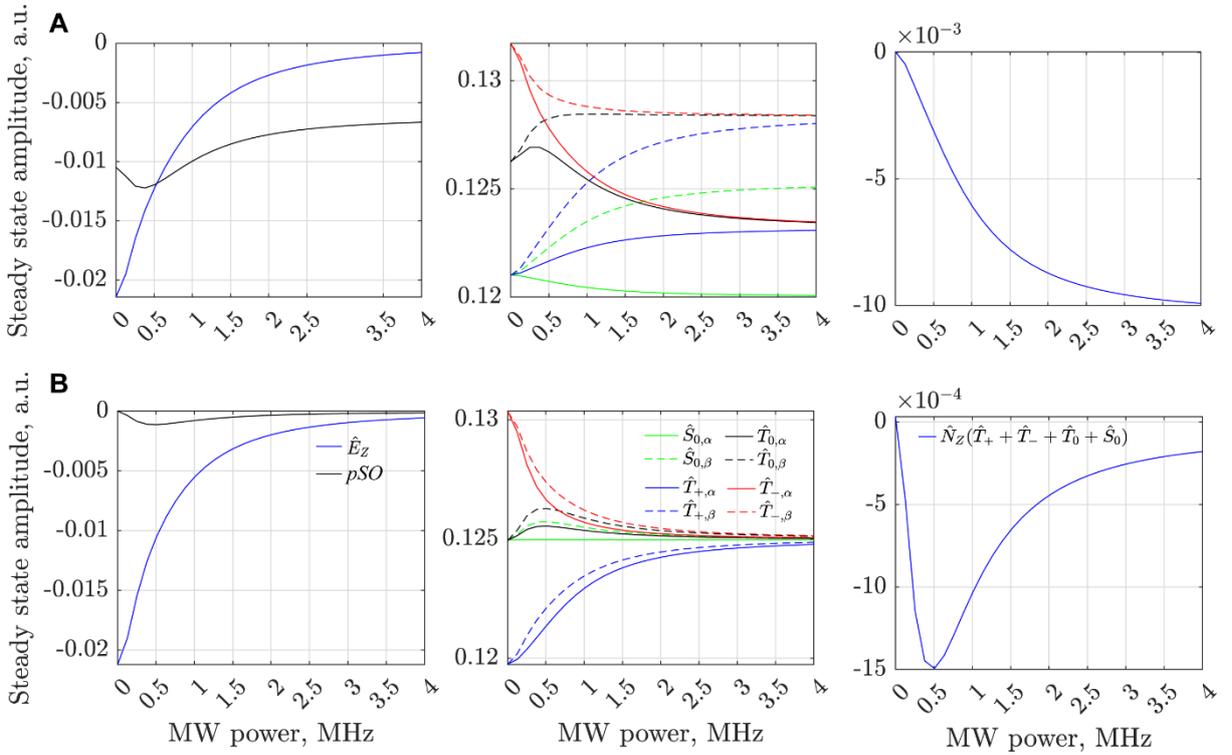

**Figure 4:** Dependence of the steady states achieved by various spin states on the microwave nutation power, for a spin system fulfilling $J_{ex} = \omega_E + \omega_N$ (A), or $J_{ex} = \omega_N$ (B). *Left-hand panels:* difference between the $\hat{T}_+$ and $\hat{T}_-$ states, corresponding to the electron longitudinal magnetization, $\hat{E}_Z$, and between the $\hat{T}_0$ and $\hat{S}_0$ states, corresponding to the pseudo singlet order $pSO = -\hat{E}_{1-}\hat{E}_{2+} - \hat{E}_{1+}\hat{E}_{2-}$. *Center panels:* amplitude of the α and β nuclear components of the singlet and triplet states. *Right-hand panels:* longitudinal nuclear magnetization $\hat{N}_Z$, arising from the sum of the $\hat{N}_Z\hat{T}_+$, $\hat{N}_Z\hat{T}_0$, $\hat{N}_Z\hat{T}_-$ and $\hat{N}_Z\hat{S}_0$ components, according to Eq. 15. All simulations were performed assuming 9.4 T using the parameters given in Table 1.

When on-resonance microwave irradiation is applied at the electron Larmor frequency of the radical, meaning $-\omega_{off} = \omega_E$, the eigenvalues in Eq. (5) become:

$$E_1(t) = -\frac{3J_{ex}}{4} + \frac{\omega_N}{2} \rightarrow |\hat{S}_{0,\alpha}\rangle \qquad E_2(t) = \frac{J_{ex}}{4} + \frac{\omega_N}{2} + \omega_{MW} \rightarrow |\hat{T}_{+,\alpha}\rangle$$
$$E_3(t) = \frac{J_{ex}}{4} + \frac{\omega_N}{2} \rightarrow |\hat{T}_{0,\alpha}\rangle \qquad E_4(t) = \frac{J_{ex}}{4} + \frac{\omega_N}{2} - \omega_{MW} \rightarrow |\hat{T}_{-,\alpha}\rangle \qquad (9)$$

corresponding to the indicated eigenstates, while the eigenvalues in Eq. (6) become:

$$E_5(t) = -\frac{3J_{ex}}{4} - \frac{\omega_N}{2} \rightarrow |\hat{S}_{0,\beta}\rangle \qquad E_6(t) = \frac{J_{ex}}{4} - \frac{\omega_N}{2} + \omega_{MW} \rightarrow |\hat{T}_{+,\beta}\rangle$$
$$E_7(t) = \frac{J_{ex}}{4} - \frac{\omega_N}{2} \rightarrow |\hat{T}_{0,\beta}\rangle \qquad E_8(t) = \frac{J_{ex}}{4} - \frac{\omega_N}{2} - \omega_{MW} \rightarrow |\hat{T}_{-,\beta}\rangle \qquad (10)$$

where the associated states are also indicated. The microwaves do not affect the populations of the $\hat{S}_{0,\alpha/\beta}$ and $\hat{T}_{0,\alpha/\beta}$ states, but they will lead to the decrease/increase in the energy of the $\hat{T}_{+,\alpha/\beta}$ and $\hat{T}_{-,\alpha/\beta}$ states, that will then approach the energy of the $\hat{T}_{0,\alpha/\beta}$ states (Figure 4). The left panels in Figure 4, show the electronic saturation as a function of the microwave power. In the presence of microwave irradiation the triplet states are brought out of the thermal equilibrium, and a population imbalance between $\hat{T}_+$ and $\hat{T}_-$ is transferred to a population imbalance between $\hat{T}_0$ and $\hat{S}_0$ (see Figure 4, left panels). In the case of $J_{ex} \approx \omega_E$, the maximum enhancement is achieved when the electron is fully saturated (meaning $\hat{E}_Z = \hat{T}_+ - \hat{T}_- = 0$, which in this case occurs with $\omega_{MW} > 4$ MHz); a maximum imbalance between the α and β nuclear components of the triplet states is then obtained, as shown the Figure 4A, central panel. In the case of $J_{ex} \approx \omega_N$, the maximum enhancement is achieved when the maximum imbalance between the between $\hat{T}_0$ and $\hat{S}_0$ states is obtained. This does not happen when the electron is fully saturated, but rather when maximal differences between the α and β nuclear components of the triplet states is present (Figure 4B, central panel). This explains the difference between the monotonic increase of the JDNP enhancement with microwave power arising in the $J_{ex}=\omega_E+\omega_N$ case, vs the maximum enhancement observed at about 0.5 MHz that decreases with further microwave power in the case of $J_{ex}=\omega_N$ (e.g., Figure 3).

In the presence of relaxation due to local vibrational modes, the self-relaxation rates of the α and β nuclear components of the singlet and of the triplet states become of comparable magnitude (see Figures S2 and S3 in the Supporting Material). Therefore, within the framework described so far, the imbalance between their populations is suppressed, and with it the nuclear polarization enhancement. The creation of a steady state DNP enhancement is thus created by a different mechanism in the presence of the local vibrational relaxation. This DNP enhancement, which was exemplified in Figures 1-3, can be explained using cross-relaxation arguments akin to those arising in ODNP. Specifically:

1) JDNP mechanism for a biradical with $J_{ex} \approx \omega_E$

Figure 5 shows the difference between the cross-relaxation rates among the $\hat{T}_{+,\beta}$ and $\hat{S}_{0,\alpha}$ and among the $\hat{T}_{+,\alpha}$ and $\hat{S}_{0,\beta}$ states, as a function of magnetic field and of $J_{ex}$ in the presence of both Redfield and non-Redfield (local) relaxation terms. Notice the marked differences that arise in these rates when the condition $J_{ex} \approx \omega_E$ is fulfilled (assuming here a negative inter-electron exchange coupling; otherwise the condition $J_{ex} \approx -\omega_E$ would have to be fulfilled). This difference arises due to the presence of different denominators in the spectral density functions describing the cross-relaxation rates, that becomes relevant when the condition $J_{ex} \approx \omega_E$ is satisfied:

$$\sigma_{\hat{T}_{+\beta},\hat{S}_{0\alpha}} = -\frac{\Delta_{HFC}^2}{180} J\left(J_{ex} + \omega_E - \omega_N\right) \tag{11}$$

$$\sigma_{\hat{T}_{+\alpha},\hat{S}_{0\beta}} = -\frac{\Delta_{HFC}^2}{30} J\left(J_{ex} + \omega_E + \omega_N\right) \tag{12}$$

Here the $\Delta_{AHF}^2$ term is the second rank norm squared [20, 43] arising from anisotropies associated to the difference between the hyperfine coupling tensors between the proton and the two electrons in the system; and $J(\omega)$ is the spectral density function at a frequency $\omega$, corresponding to $\tau_C/(1+\tau_C^2 \omega^2)$. A similar discussion can be made also in the case of a positive $J_{ex} \approx -\omega_E$, leading to $\sigma_{\hat{T}_{-\beta},\hat{S}_{0\alpha}} > \sigma_{\hat{T}_{-\alpha},\hat{S}_{0\beta}}$ at any magnetic field.

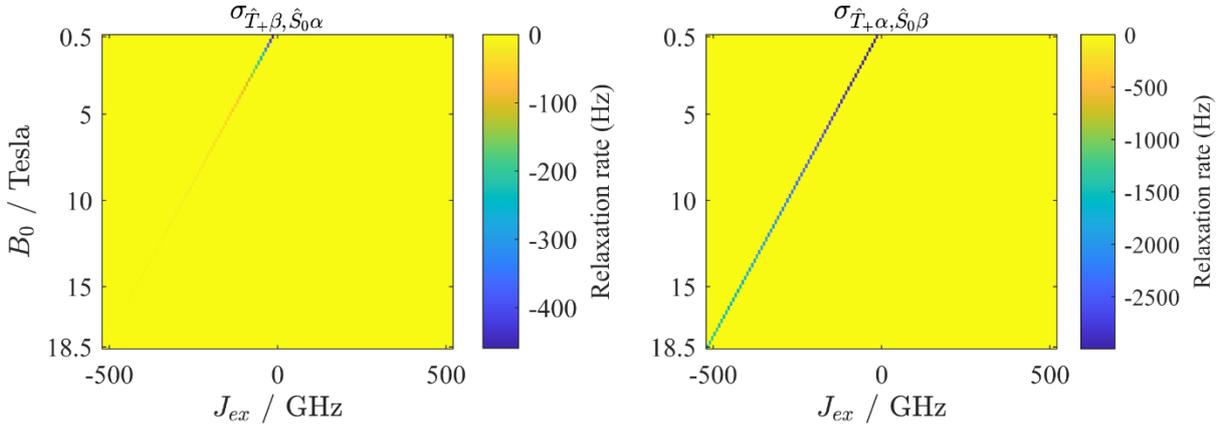

**Figure 5:** Triplet to singlet cross-relaxation rates $\sigma_{\hat{T}_{+\beta},\hat{S}_{0\alpha}}$ and $\sigma_{\hat{T}_{+\alpha},\hat{S}_{0\beta}}$ as a function of the $J_{ex}$ ranging between -500 GHz and +500 GHz and of the magnetic field. Simulation parameters are given in the Table 1.

2) JDNP mechanism for a biradical with $J_{ex} \approx \omega_N$

Figure 6 shows the difference between the cross-relaxation rates among $\hat{S}_{0,\alpha}$ and $\hat{T}_{0,\beta}$ states and among the $\hat{S}_{0,\beta}$ and $\hat{T}_{0,\alpha}$ states with the magnetic field and with $J_{ex}$, that occurs when the condition $J_{ex} \approx \omega_N$ is fulfilled. The difference between the triplet to singlet cross-relaxation rates arises once again due to the presence of $(J_{ex} - \omega_N)$ and $(J_{ex} + \omega_N)$ in the denominators of the spectral density functions describing the rates:

$$\sigma_{\hat{T}_{0\beta},\hat{S}_{0\alpha}} = -\frac{\Delta_{HFC}^2}{60} J\left(J_{ex} - \omega_N\right) \tag{13}$$

$$\sigma_{\hat{T}_{0\alpha},\hat{S}_{0\beta}} = -\frac{\Delta_{HFC}^2}{60} J\left(J_{ex} + \omega_N\right) \tag{14}$$

This difference becomes maximum when the condition $J_{ex} = \omega_N$ is fulfilled.

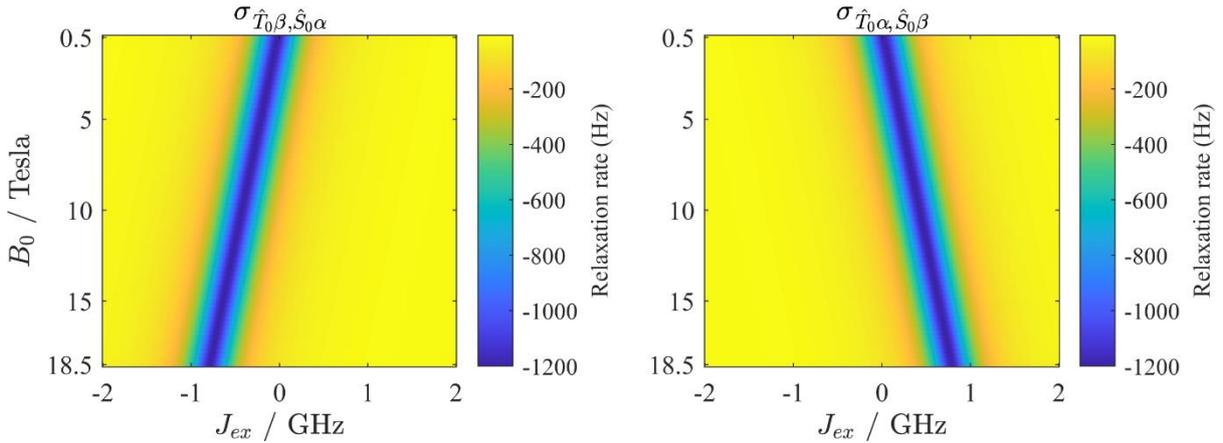

**Figure 6:** Triplet to singlet cross-relaxation rates $\sigma_{\hat{T}_{0\beta},\hat{S}_{0\alpha}}$ and $\sigma_{\hat{T}_{0\alpha}\hat{S}_{0\beta}}$, as a function of the $J_{ex}$ and of the magnetic field. The simulation parameters are given in the Table 1.

### 4. Discussion and Conclusions

The present study revisited the proposal for enhancing NMR signals in solution state at any magnetic field and for a wide range of rotational correlation times, solely relying on inter-molecular dipolar hyperfine couplings and the presence of polarizing biradicals with either $J_{ex} \approx \omega_E$ or $J_{ex} \approx \omega_N$. In the previous study, which neglected relaxation from local vibrational modes, the JDNP condition led to a transient imbalance between the α and β nuclear components of singlet and triplet state populations, and consequently to a transient nuclear magnetization build-up. The addition of relaxation from local vibrational modes to the Redfield relaxation superoperator, suppresses the transient nature of this imbalance and leads to a steady state enhancement, akin to that arising in ODNP. However, unlike steady state ODNP effects whose enabling cross-relaxation rates decay quadratically with the strength of the magnetic field, the JDNP effects are only weakly dependent on magnetic fields. Thus, if the available microwave nutation powers are sufficiently high to compete with the triplets' self-relaxation rates (~$10^5$Hz), a cross-relaxation process between the out-of-the equilibrium triplet states – either the $\hat{T}_{\pm}$ in the $J_{ex} \approx \omega_E$ case or the $\hat{T}_0$ in the $J_{ex} \approx \omega_N$ case – and the singlet state, will take place. When either the conditions $J_{ex} \approx \pm\omega_E$ or $J_{ex} \approx \pm\omega_N$ are fulfilled, these triplet-to-singlet cross-relaxation rates will occur with different rates for α and β states (Eqs. 11 – 12 and Eqs. 13 – 14). This leads to an imbalance between the populations of α and β nuclear components, and consequently to the creation of longitudinal nuclear polarization in according to:

$$\hat{N}_Z = \left(\hat{T}_{+,\alpha} - \hat{T}_{+,\beta}\right) + \left(\hat{T}_{0,\alpha} - \hat{T}_{0,\beta}\right) + \left(\hat{T}_{-,\alpha} - \hat{T}_{-,\beta}\right) + \left(\hat{S}_{0,\alpha} - \hat{S}_{0,\beta}\right)$$
$$= \hat{N}_Z\hat{T}_+ + \hat{N}_Z\hat{T}_0 + \hat{N}_Z\hat{T}_- + \hat{N}_Z\hat{S}_0 \qquad (15)$$

These differences are the ones shown on the right-hand side of Figure 4, evidencing the enhancement of the nuclear polarization. Notice that due to the cancellation of the electron or the nuclear Larmor frequencies in the dominators of the spectral density functions arising under JDNP, these cross-relaxation rates are solely dependent on dipolar hyperfine interactions; hence the weak field dependencies shown in Figure 2. Notice as well that while in the $J_{ex} \approx \omega_E$ case the sign of the enhancement is always negative, for $J_{ex} \approx \omega_N$ the sign of the enhancement will change together with the sign of the $J_{ex}$ (see Supporting Material for more details).

The present study assumed a fixed electrons-nuclear geometry. In realistic cases, the nuclear polarization build-up will be interrupted by diffusion of the proton being enhanced, out of the polarization region where JDNP is active and into non-enhancing regions. Under continuous microwave irradiation, however, a steady state enhancement is still expected to arise, as the proton diffuses from one polarizing environment into another in the solution; this is further discussed in the Supporting material.

Another issue further discussed in the Supporting material is the slight increase in the JDNP enhancement with rotational correlation time shown in Figure 3. This reflects the increased differences between the α/β and the β/α cross-relaxation rates with magnetic field, as expected from the spectral density changes at the JDNP conditions. Notice, however, that in actual cases singlet and triplet self-relaxation rates will also depend on the $g$-anisotropy, [20, 22] and that a large difference between these would lead to a decrease of the triplet $T_1(s)$, subtracting efficiency from JDNP. Therefore, the ideal biradicals for these JDNP experiments would be those connecting two identical monomers, with axially symmetric $g$-tensors; these are often found in bistrityls,[33] with a linear, short, conformationally rigid linker. Inter-electron exchange couplings in the order of the electron Larmor frequency can be achieved by linking these radicals with a single para-phenylene unit, leading to a $J_{ex}$ of ≈240 ±25 GHz [36]. Inter-electron exchange couplings in the order of the nuclear Larmor frequency can be achieved by lengthening the linker. [37] By contrast, nitroxide biradicals are not suitable for JDNP due to their large $g$-tensor anisotropy that increases with the magnetic field, contributing to shortening the electron's $T_1$ and subtracting efficiency to the JDNP process. Pure hydrocarbon-based biradicals, characterized by axial $g$-tensors with less anisotropy and characterized by slower electron relaxation rates [39], are expected to be the most promising polarizing agents for JDNP. Electron Paramagnetic Resonance (EPR) / DNP instrumentation operating at high magnetic field [12, 38, 44] and equipped with suitable microwave power sources, [38] could enable the observation of the JDNP and development of the optimal polarizing agents.

## 5. Acknowledgments

This project was funded by the Israel Science Foundation (ISF 1874/22), the Minerva Foundation (Germany), and the US National Science Foundation (grants number CHE-2203405). MGC acknowledges Weizmann's Faculty of Chemistry for a Dean Fellowship. LF holds the Bertha and Isadore Gudelsky Professorial Chair and Heads the Clore Institute for High-Field Magnetic Resonance Imaging and Spectroscopy whose support is acknowledged, as is the generosity of the Perlman Family Foundation. The authors acknowledge Prof. Ilya Kuprov, Prof. Olav Schiemann and Mr Kevin Kopp for discussions.

**Supplementary material for**

**Steady state effects introduced by local relaxation modes on J-driven DNP-enhanced NMR**

Maria Grazia Concilio* and Lucio Frydman

*Department of Chemical and Biological Physics, Weizmann Institute of Science, Rehovot, Israel*

**Contents**

A. Effect of *g*-tensors' relative orientation on the DNP enhancement
B. Evolution of the cross-relaxation rate $\hat{E}_z \rightarrow \hat{N}_z$ with the magnetic field in a biradical spin system
C. Effect of the relaxation arising from local vibrations on the self-relaxation rates of the α and β nuclear components of singlet and triplet states
D. Cross-relaxation rates $\sigma_{\hat{T}_{+\alpha},\hat{S}_{0\beta}}$ and $\sigma_{\hat{T}_{+\beta},\hat{S}_{0\alpha}}$ as a function of the shift from the condition $J_{ex}=\omega_E$
E. JDNP enhancement in the space surrounding the biradical
F. Effect of the proton diffusion on the DNP enhancement
G. Effect of the rotational correlation time on the singlet – triplet cross-relaxation rates

*Email*: maria-grazia.concilio@weizmann.ac.il

## A. Effect of *g*-tensors' relative orientation on the DNP enhancement

In solution at room temperature, the rotational barrier for mono para-phenyl and acetylene linked biradicals is of about 1 - 10 kJ/mol, resulting in a free internal rotation. [37, 45] Therefore, *g*-tensors are likely to be non-colinear, their rotation about the trityl's axis can be described using a directional cosine matrix, in according to the active ZYZ Euler rotation convention $VDV^{-1}$, where V is the rotational matrices and D is the interaction tensor, see Figure S1. [46]

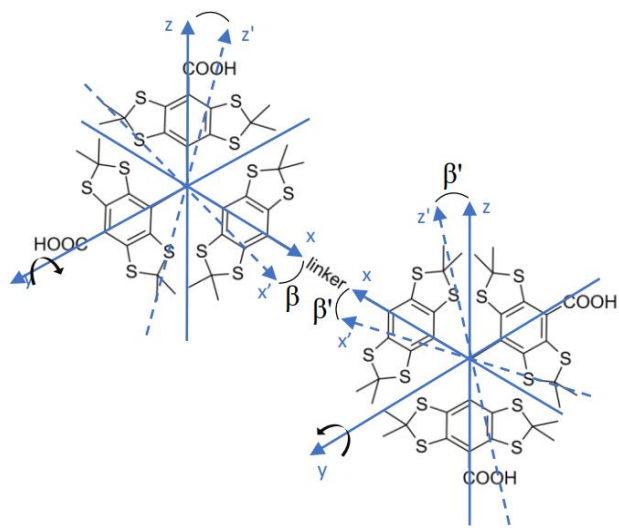

**Figure S1:** Molecular structure of a trityl biradical, showing the pseudo-$C_3$ symmetry axis of each mono-radical units [47] and the *g*-tensor rotation about the Y-axis leading to the β (β') angle.

Due to the trityl's *g*-tensor axial symmetry, [33] only rotations about the Y-axis, on the linker connecting the two monomeric units in a symmetric biradical, can lead to variation of the *g*-tensors, and a difference between the two anisotropic *g*-tensors, that will rob efficiency from JDNP, as already discussed in [22, 48] and shown in Figures S2. Rotations about the X- and Y-axis lead to no changes in the DNP enhancement. As discussed in the main text, the β bending angle in long (~ 38 Å) linear linkers containing phenylene and acetylene units ranges between about 0° and ±40°, [35] smaller displacements were expected for shorter linkers (~14 Å – 20 Å), considered for JDNP at 9.4 T. However, differences between two anisotropic *g*-tensors will only lead to electron relaxation and not to the JDNP.

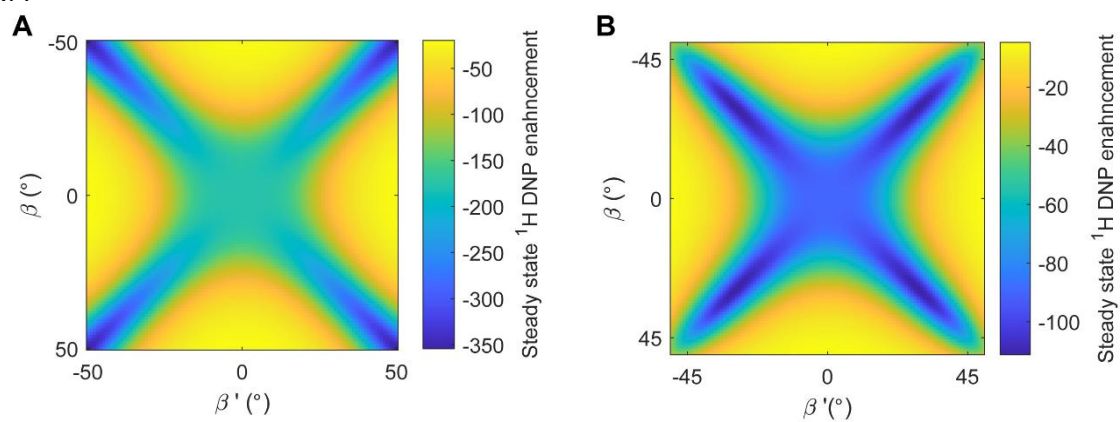

**Figure S2:** $^1$H JDNP enhancements calculated at 9.4 T using a microwave nutation power of 500 kHz as a function of the Euler angles β and β' for the *g*-tensor, using the trityl coordinate system shown in Figure S1. (A) $J_{ex} = \omega_E + \omega_N$. (B) $J_{ex} = \omega_N$. The β and β' angles were varied between 0° and ±45°, while the angles α (α') and γ (γ') were set to zero.

At 9.4 T, the average enhancement obtained over all the β angles, shown in Figure S2, is equal to -125 for $J_{ex} = \omega_E + \omega_N$ and to -49 for $J_{ex} = \omega_N$ respectively. We performed the same analysis at the magnetic field of 0.5 T, 3.4 T, 14.1 T and 18.5 T, the average enhancements are equal to -177, -153, -90 and -68 respectively for $J_{ex} = \omega_E + \omega_N$ and equal to 1, -16, 42 and -32 for $J_{ex} = \omega_N$. These values are comparable to those shown in Figure 2 in the main text, obtained fixing the β equal to π/8.

## B. Evolution of the cross-relaxation rate $\hat{E}_Z \to \hat{N}_Z$ with the magnetic field in a biradical spin system

In the case of a biradical with $J_{ex} = \omega_E + \omega_N$, it is possible to observe a non-zero cross-relaxation rate $\hat{E}_Z \to \hat{N}_Z$ at any magnetic field, while this rate is zero for a biradical with $J_{ex} = \omega_N$, see Figure S1. This is because of the absence of the $\hat{E}_Z$ state in the $\hat{T}_0$ and $\hat{S}_0$ states, involved in the JDNP polarization mechanism for a biradical with $J_{ex} = \omega_N$. [20] By contrast, non-zero cross-relaxation rates between $\hat{N}_Z$ and both the singlet order $SO$ and the pseudo singlet order $pSO$, are present at any magnetic field for both the conditions $J_{ex} = \omega_E + \omega_N$ and $J_{ex} = \omega_N$ (Figure S3). The $SO$ and $pSO$ singlet and pseudo-singlet order are quantities arising from the imbalance between the populations of the three triplet states and the singlet state, and from the imbalance between the $\hat{T}_0$ and $\hat{S}_0$ states, respectively. Singlet and pseudo-singlet order are thus dictated by the inter-electron exchange coupling, and correspond to the expectation values of the operators $-\frac{2}{3}\left(\hat{E}_{1-}\hat{E}_{2+} + \hat{E}_{1+}\hat{E}_{2-} + 2\hat{E}_{1Z}\hat{E}_{2Z}\right)$ and $-\hat{E}_{1-}\hat{E}_{2+} - \hat{E}_{1+}\hat{E}_{2-}$, respectively. Notice also that the self-relaxation rates of the states $-R[pSO]$, $-R[SO]$ and $-R[\hat{E}_Z]$ increase with increasing magnetic fields, while the rate $-R[\hat{E}_Z]$ remains constant with magnetic field for a biradical with $J_{ex} = \omega_N$, as it was observed for a monoradical. [20]

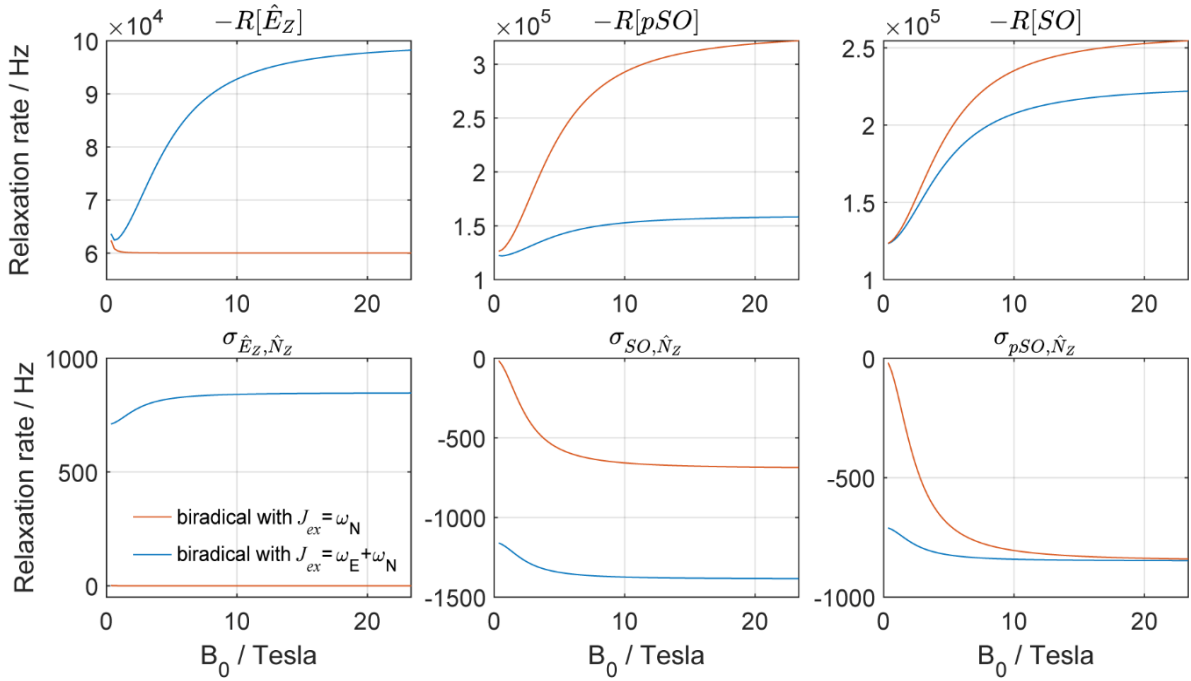

**Figure S3:** Self-relaxation rates $-R[\hat{E}_Z]$, $-R[pSO]$ and $-R[SO]$, and cross-relaxation rates $\sigma_{\hat{E}_Z,\hat{N}_Z}$, $\sigma_{pSO,\hat{N}_Z}$ and $\sigma_{SO,\hat{N}_Z}$ as a function of the magnetic field for a biradical with the parameters shown in the Table 1.

## C. Effect of the relaxation arising from local vibrational modes on the self-relaxation rates of the α and β nuclear components of the singlet and triplet states

Numerical self-relaxation rates predicted by the Redfield model [49] as a function of the magnetic field of the $J_{ex}$ were already reported in Ref [20]. When a relaxation arising from molecular vibrations, possessing for instance a 6x10$^4$ Hz rate as observed in trityl radicals [24, 28], is applied to the longitudinal and transverse terms in the Redfield relaxation superoperator, [50] the Redfield-predicted rates increase by several orders of magnitude. The difference between α and β nuclear components of singlet and triplet states becomes negligible, also at the conditions $J_{ex}=\omega_E+\omega_N$ and $J_{ex}=\omega_N$, see Figures S4 and S5 respectively. This would result in the suppression of the transient JDNP described previously, but lead to the steady state enhancement shown in Figures 1-3 of the main text.

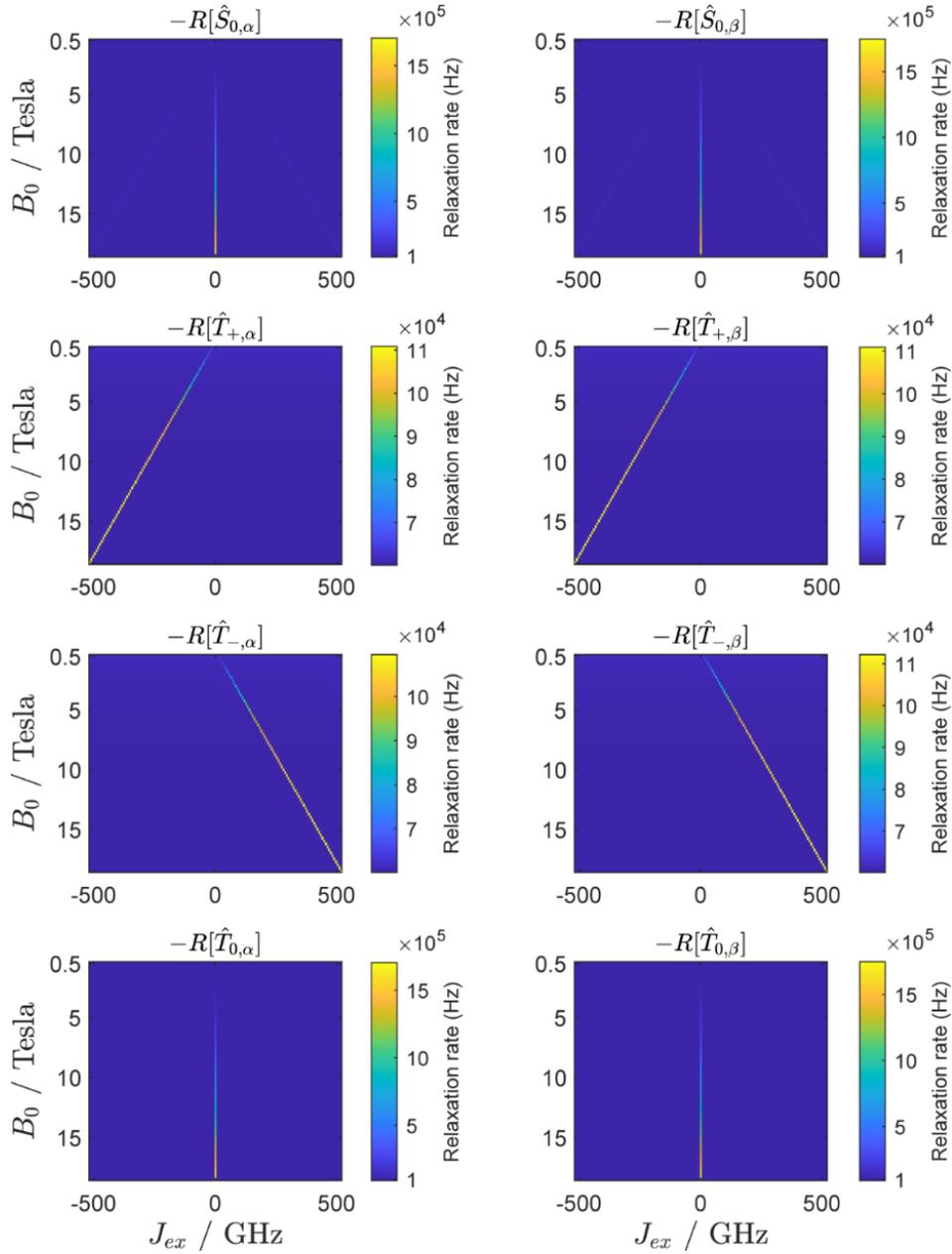

**Figure S4:** Self-relaxation rates of the α and β nuclear components of the singlet and of the triplet states as a function of a $J_{ex}$ changing between – 500 GHz and 500 GHz and of the magnetic field, $B_0$. The simulation parameters are given in the Table 1 of the main text.

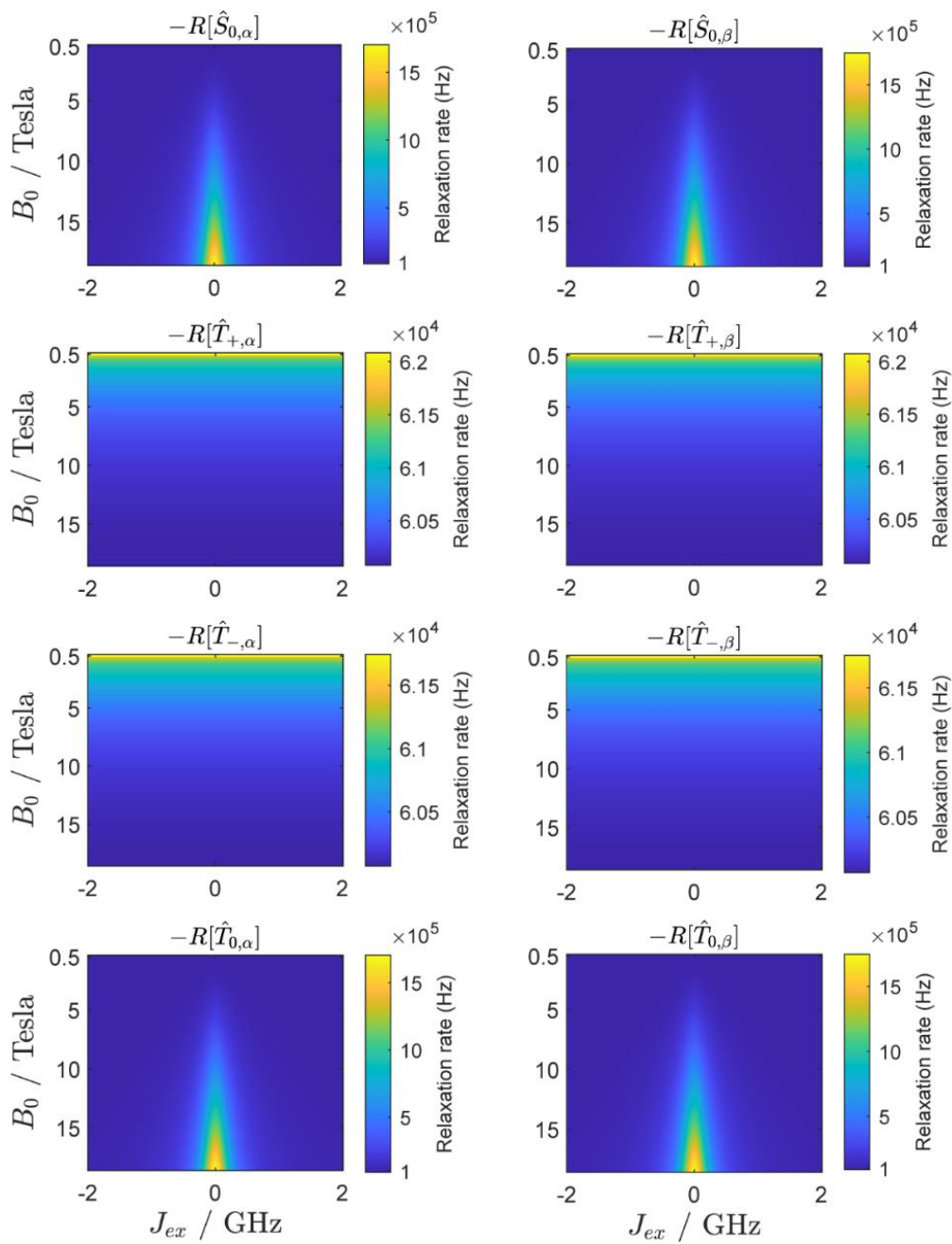

**Figure S5:** Self-relaxation rates of the α and β nuclear components of the singlet and of the triplet states as a function of a $J_{ex}$ changing between – 2000 MHz and 2000 MHz and of the magnetic field, $B_0$. The simulation parameters are given in the Table 1 of the main text.

Relaxation rates shown in Figure S4 and S5, dominated by relaxation arising from local vibrational modes, correspond to $T_1(s)$ of about 10 µs (see Figure S6 and S7); consistent with values found in the literature for trityl radicals. [24, 25, 28] The local mode relaxation affects mostly the longitudinal relaxation terms, and has a minor effect on the transverse relaxation, whose rates corresponds to $T_2(s)$ ranging between 10 and 100 ns, and are dominated by inter-electron dipolar interactions. [20]

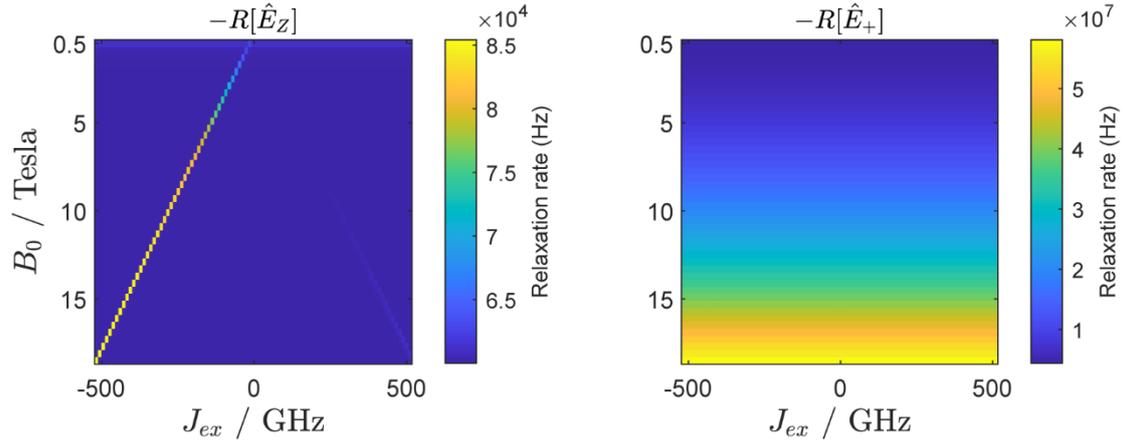

**Figure S6:** Longitudinal and transverse electron relaxation as a function of a $J_{ex}$ changing between – 500 GHz and 500 GHz and of the magnetic field, $B_0$. The simulation parameters are given in the Table 1 of the main text.

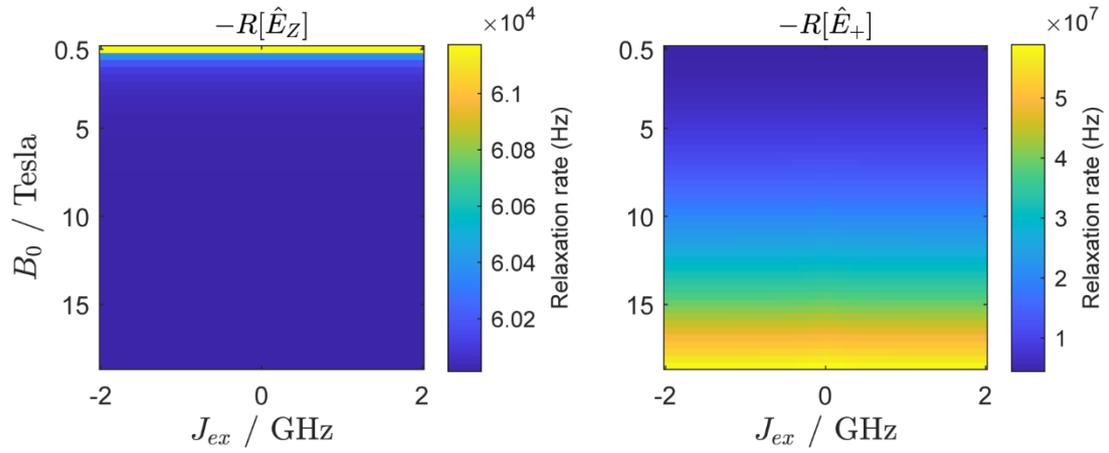

**Figure S7:** Longitudinal and transverse electron relaxation as a function of a $J_{ex}$ changing between – 2000 MHz and 2000 MHz and of the magnetic field, $B_0$. The simulation parameters are given in the Table 1 of the main text.

### D. Cross-relaxation rates $\sigma_{\hat{T}_{+\alpha},\hat{S}_{0\beta}}$ and $\sigma_{\hat{T}_{+\beta},\hat{S}_{0\alpha}}$ as a function of shifts from the condition $J_{ex}=\omega_E$

The sign of $J_{ex}$ defines the energy difference between the singlet and the triplet manifold, [51] and its magnitude might change in solution due to variations in the inter-electron distance arising due to molecular collisions. Figure S8 shows how, as magnetic fields increase while in the neighborhood of the $J_{ex}=\omega_E$ condition, the magnitudes of the cross-relaxation rates $\sigma_{\hat{T}_{+\alpha},\hat{S}_{0\beta}}$ and $\sigma_{\hat{T}_{+\alpha},\hat{S}_{0\beta}}$ defining the JDNP enhancement, can cross one another with slight variations of $J_{ex}$. For example the case in Figure S8 shows that at magnetic fields ≥5.4 T, the cross-relaxation rate $\sigma_{\hat{T}_{+\alpha},\hat{S}_{0\beta}}$ is higher than the cross-relaxation rate $\sigma_{\hat{T}_{+\beta},\hat{S}_{0\alpha}}$ for negative $J_{ex}$(s) ranging between $\omega_E$ and $\omega_E$ + 1 GHz, while the opposite happens in a smaller range of conditions between $\omega_E$ - $\omega_N$ - 200 MHz and $\omega_E$ - $\omega_N$ + 200 MHz.

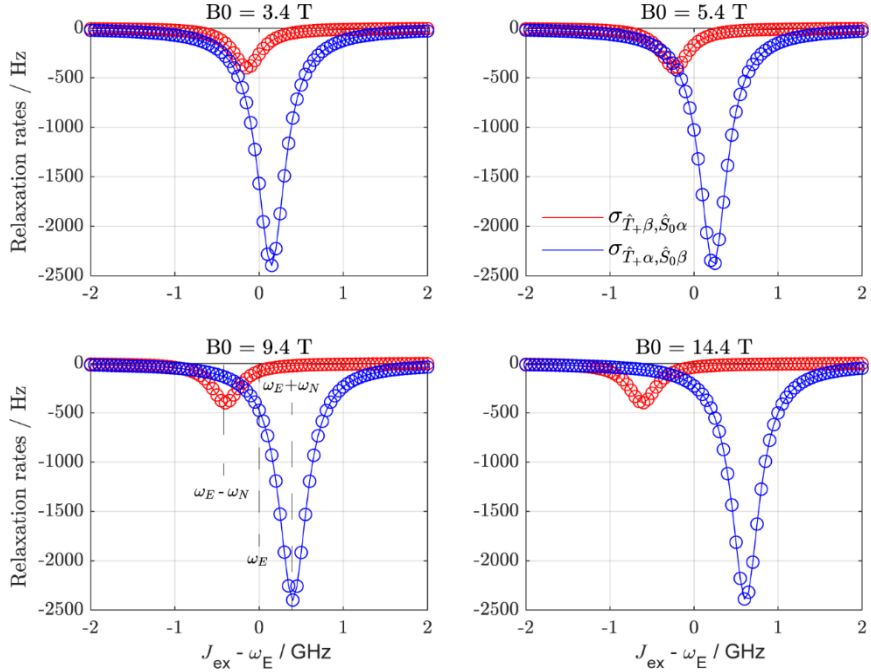

**Figure S8:** Numerical (straight lines) and analytical (circles, computed in according to Eq. 11 and Eq. 12 in the main text) triplet to singlet cross-relaxation rates $\sigma_{\hat{T}_{+\beta},\hat{S}_{0\alpha}}$ and $\sigma_{\hat{T}_{+\alpha},\hat{S}_{0\beta}}$ as a function of the $J_{ex}$ - $\omega_E$ that was varied between $\omega_E$ ± 2 GHz, at different magnetic fields. Labels in the left bottom panel indicate the conditions $J_{ex} = \omega_E - \omega_N$, $J_{ex} = \omega_E$ and $J_{ex} = \omega_E + \omega_N$. The simulation parameters are given in the Table 1 in the main text.

In all cases, cross-relaxation rates $\sigma_{\hat{T}_{+\alpha},\hat{S}_{0\beta}}$ and $\sigma_{\hat{T}_{+\beta},\hat{S}_{0\alpha}}$ will reach their maxima at $\omega_E+\omega_N$ and $\omega_E-\omega_N$ respectively, and that the rate $\sigma_{\hat{T}_{+\alpha},\hat{S}_{0\beta}}$ is higher than the rate $\sigma_{\hat{T}_{+\beta},\hat{S}_{0\alpha}}$ due to the presence of different denominators in the spectral density functions describing the cross-relaxation rates in Eq. 11 and Eq. 12 in the main text. There are no computational methods to determine the exact values of $J_{ex}$ and their distribution in solution. The latter might arise due to fluctuations in the inter-electron distance due to intra-molecular vibrations occurring in the fs - ps timescale, or as a result of slower solvent-driven molecular collisions. In either case, the predominance of the negative enhancement (Figure S8) ensures that, even in the presence values of rapidly or slowly fluctuating $J_{ex}$ values, the resultant will always be that a field can be found where there is a sizable net enhancement of the nuclear polarization.

### E. JDNP enhancement in the space surrounding the biradical

Negative and positive nuclear magnetization enhancements will be observed in the case of $\sigma_{\hat{T}_{+\alpha},\hat{S}_{0\beta}} > \sigma_{\hat{T}_{+\beta},\hat{S}_{0\alpha}}$ and $\sigma_{\hat{T}_{+\beta},\hat{S}_{0\alpha}} > \sigma_{\hat{T}_{+\alpha},\hat{S}_{0\beta}}$ respectively. Figure S9 shows the distribution of the enhancements in the space surrounding a model biradical, at the conditions $J_{ex} = \omega_E - \omega_N$, $\omega_E$ and $\omega_E + \omega_N$. Notice that the nuclear enhancement will be negative at the conditions $J_{ex} = \omega_E$ and $J_{ex} = \omega_E + \omega_N$ (except in few regions very close to the electron that are unreachable by the solvent), and positive at the condition $J_{ex} = \omega_E - \omega_N$. The negative enhancement observed at $J_{ex} = \omega_E + \omega_N$ is stronger than the positive enhancement observed at $J_{ex} = \omega_E - \omega_N$ due to the higher difference between the cross-relaxation rates at $J_{ex} = \omega_E + \omega_N$. The sign of the enhancement in space is important to ensure that the spatial averaging due to molecular translations does not lead to overall zero nuclear magnetization in the bulk, and that fluctuations in $J_{ex}$ do not average out the overall JDNP effect.

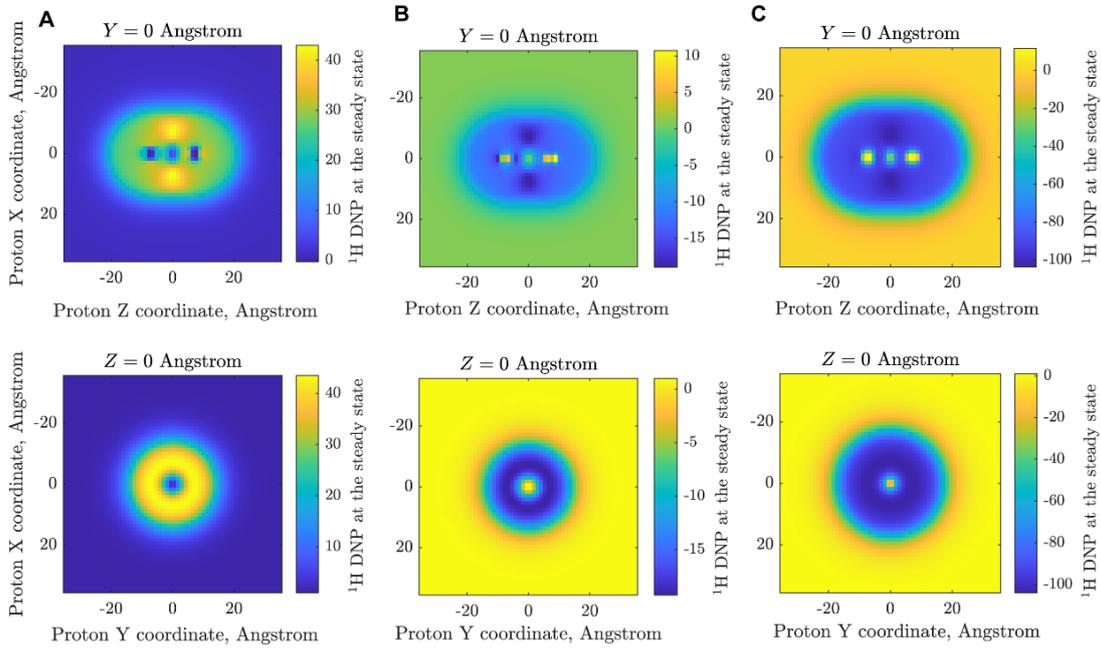

**Figure S9:** Steady DNP enhancement as a function of random $^1$H-coordinates surrounding a model biradical, calculated for 9.4 T assuming a continuous microwave irradiation with power equal to 500 kHz. The inter-electron exchange coupling was set to $\omega_E - \omega_N$ in panel A, to $\omega_E$ in panel B, and to $J_{ex} = \omega_E + \omega_N$ in panel C.

In the case of $J_{ex} = \omega_N$, the sign of the enhancement remains constant in any region in the space surrounding the biradical, see Figure S10.

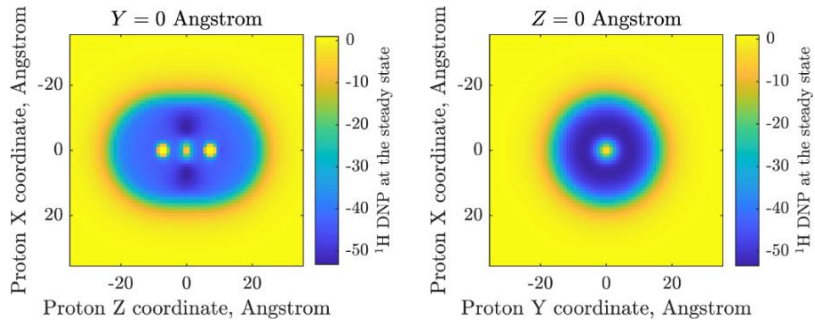

**Figure S10:** Steady DNP enhancement as a function of random 1H-coordinates surrounding a model biradical calculated at 9.4 T, using continuous microwave irradiation with power equal to 500 kHz, the inter-electron exchange coupling was set to $\omega_N$. The other simulation parameters were given in the Table 1.

For both the conditions $J_{ex} \approx \omega_E$ and $J_{ex} = \omega_N$, the polarization region where the JDNP is active, extended up to about 20 Å from both electrons. The enhancement becomes equal to zero over 20 Å away from the electrons, and when the proton is placed symmetrically in the biradical's centre (*i.e.* at [0 0 0] Å).

### F. Effect of the proton diffusion on the DNP enhancement

To take into account the constant interruption of the polarization process due to the diffusion of a solvent proton between polarizing and non-polarizing regions within a solution, a simulation scheme was set up where proton coordinates were constantly exchanged between regions close and far from the electrons. As mentioned in previous work, [52] where we performed random walk analysis of the proton diffusion in a solution with ≈10 mM biradical concentration, we determined that the mean occurrence of a solvent proton will be about 10% within the polarizing region and the remaining 90% in an inter-biradical, non-polarizing region. Considering then that the proton will be diffusing randomly with diffusion constant of ≈1 µm$^2$/ms (*i.e.* ≈10$^{-9}$ m$^2$ s$^{-1}$ diffusivity constant at 25 °C [4]), a time-stepped simulation was set where the proton was allowed to vary randomly in 1x10$^{-6}$ s and 9x10$^{-6}$ s steps for the polarizing and non-polarizing regions, respectively. Figure S11 shows the nuclear polarization build-up obtained in the time for both the conditions $J_{ex} = \omega_E + \omega_N$ and to $J_{ex} = \omega_N$, with and without considering the diffusion process. Notice that the nuclear polarization build-up reaches steady state after 10(s) of ms, but there is no decrease of steady state enhancement. This model reproduced the actual experiment, where the proton will transiently interact but then leave one biradical, then form another transient proton - biradical complex and become slightly more polarized, and so on, under continuous microwave irradiation, till a steady state enhancement is reached.

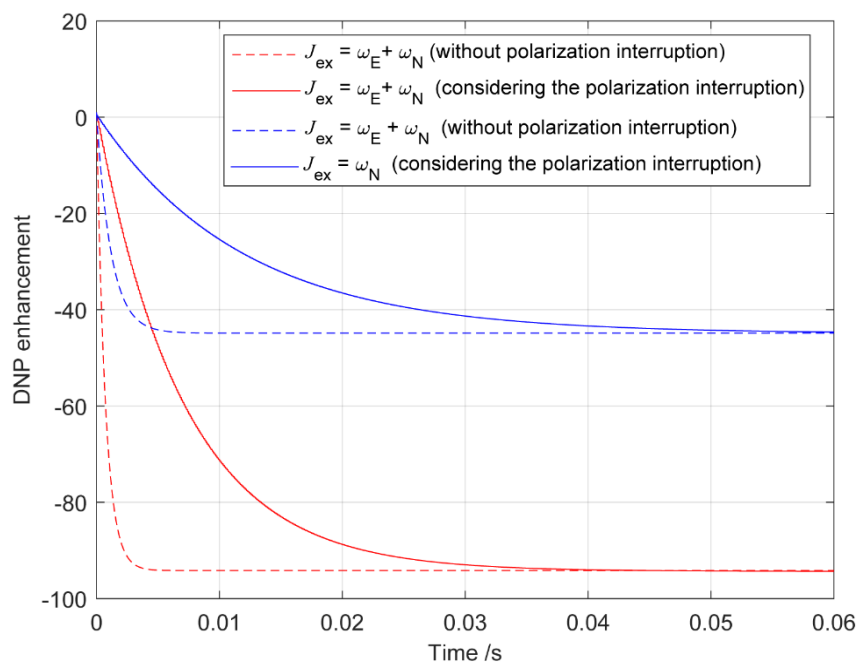

**Figure S11:** Time-domain simulations performed as described in Refs. [20] and [8], assuming a three-spin system where the solvent proton exchanges between configurations that are inside and outside a polarization region. The coordinates of the proton close to the electrons were kept the same as those given in the Table 1, while the coordinates of the electron far away were set to [0 20 5]. In the latter, distances between the two unpaired spin-1/2 electrons (belonging to the radical) and the spin-1/2 proton (assumed to belong to a stationary solvent molecule) were set to 20 Å and 23 Å respectively. Simulations were performed at 9.4 T, using continuous microwave irradiation with power equal to 500 kHz; the $J_{ex}$ was set to $\omega_E + \omega_N$ (blue line) and to $\omega_N$ (red line).

## G. Effect of the rotational correlation time on the singlet – triplet cross-relaxation rates

Figure S12 shows that a difference between the cross-relaxation rates $\sigma_{\hat{S}_{0\alpha},\hat{T}_{+\beta}}$ and $\sigma_{\hat{S}_{0\beta},\hat{T}_{+\alpha}}$ (or $\sigma_{\hat{S}_{0\alpha},\hat{T}_{0\beta}}$ and $\sigma_{\hat{S}_{0\beta},\hat{T}_{0\alpha}}$) will be present at any $B_0$, and will increase with the correlation time $\tau_C$. Spectral density functions describing the cross-relaxation rates (Eqs. 11-14 in the main text) have the square of the rotational correlation time in their denominator, and as it increases so do their difference, explaining the increase of the DNP enhancement with the $\tau_C$ as shown in Figure 3 of the main text.

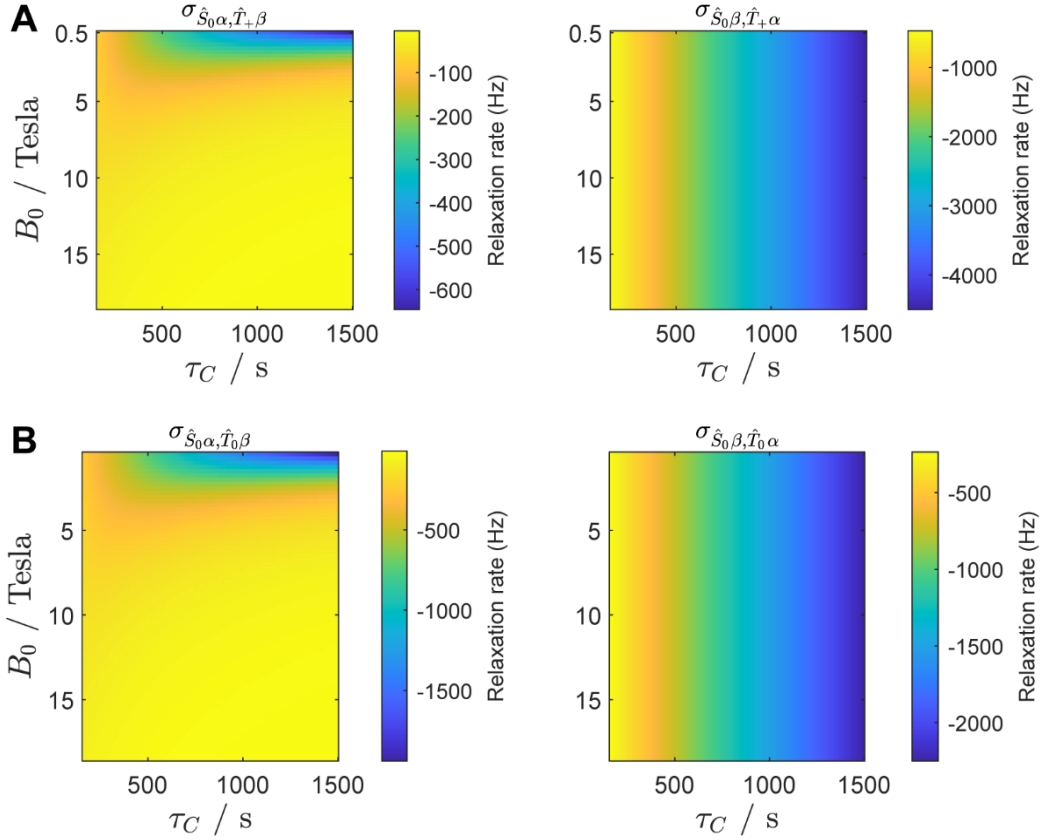

**Figure S12:** Triplet to singlet cross-relaxation rates $\sigma_{\hat{S}_{0\alpha},\hat{T}_{+\beta}}$ and $\sigma_{\hat{S}_{0\beta},\hat{T}_{+\alpha}}$ (A), and cross-relaxation rates $\sigma_{\hat{S}_{0\alpha},\hat{T}_{0\beta}}$ and $\sigma_{\hat{S}_{0\beta},\hat{T}_{0\alpha}}$ (B) as a function of the rotational correlation time $\tau_C$ and of the $B_0$. The $J_{ex}$ was set at $\omega_E+\omega_N$ in (A), and equal to $\omega_N$ in (B). The simualtion parameters are given in the Table 1 in the main text.